\documentclass[%
 aip,
cp,  
 amsmath,amssymb,
 reprint,%
]{revtex4-2}

\usepackage{graphicx}
\usepackage{dcolumn}
\usepackage{bm}
\usepackage{xcolor}

\usepackage[utf8]{inputenc}
\usepackage[T1]{fontenc}
\usepackage{mathptmx} 

\begin{document}

\title{Constraining Neutrinoless Double-Beta Decay Matrix Elements from Ab Initio Nuclear Theory}

\author{A. Belley}
 \email[Corresponding author: ]{abelley@triumf.ca}
 \affiliation{
  TRIUMF, 4004 Wesbrook Mall, Vancouver BC V6T 2A3, Canada
  }
  \affiliation{
  Department of Physics \& Astronomy, University of British Columbia, Vancouver, British Columbia V6T 1Z1, Canada
  }
  
\author{T. Miyagi}%
  \affiliation{
  Technische Universit\"at Darmstadt, Department of Physics, 64289 Darmstadt, Germany
  }
  \affiliation{
  ExtreMe Matter Institute EMMI, GSI Helmholtzzentrum f\"ur Schwerionenforschung GmbH, 64291 Darmstadt, Germany
  }
  
\author{S. R. Stroberg}%
  \altaffiliation[Current address:]{
  Department of Physics and Astronomy,
  University of Notre Dame,
  Notre Dame, IN, 46556, USA
  }
  \affiliation{
  Physics Division, Argonne National Laboratory, Lemont, IL, 60439, USA
  }

\author{J. D. Holt}%
  \affiliation{
  TRIUMF, 4004 Wesbrook Mall, Vancouver BC V6T 2A3, Canada
  }%
  \affiliation{
  Department of Physics, McGill University, Montr\'eal, QC H3A 2T8, Canada
  }
  
\date{\today}

\begin{abstract}
As experimental searches for neutrinoless double-beta ($0\nu\beta\beta$) decay are entering a new generation, with hopes to completely probe the inverted mass hierarchy, the need for reliable nuclear matrix elements, which govern the rate of this decay, is stronger than ever. 
Since a large discrepancy in results is typically found with nuclear models \cite{Engel2017}, a large unknown still exists on the sensitivity of these experiments to the effective neutrino mass. 
We consider this problem from a first-principles perspective, using the ab initio valence-space in medium similarity renormalization group.
In particular, we study correlations of the $0\nu\beta\beta$-decay matrix elements in $^{76}$Ge with other observables, such as the double Gamow-Teller giant resonance, from 34 input chiral interactions in an attempt to constrain our uncertainties and investigate the interaction dependence of the nuclear matrix element.
\end{abstract}

\maketitle

\section{\label{sec:intro}Introduction}

The observation of neutrinoless double-beta ($0\nu\beta\beta$) decay, a hypothetical process in which two neutrons transform into two protons without emitting an antineutrino, would show that lepton number is not a conserved quantity, thus having important implications for the matter-antimatter asymmetry puzzle~\cite{Fukugita1986}. 
As the nuclear matrix element (NME) of this decay is intrinsically related to its half-life, a reliable way of obtaining the NME of this transition with associated theoretical uncertainty is of crucial importance in order to compare experimental limits in different isotopes, pinpoint the mechanism responsible for lepton number violation, and extract key quantities such as the absolute mass scale of the neutrino upon potential observation of the process~\cite{Engel2017,Helo2013,Cirigliano2022}.
Ab initio nuclear theory offers an unparalleled tool to tackle the challenging task of computing the nuclear matrix elements, as it is based on systematically improvable methods that allow for rigorous uncertainty quantification~\cite{Hergert2020}. 
In these first-principle methods, the nuclear Hamiltonian is constructed from two- (NN) and three-nucleon (3N) forces obtained from chiral effective field theory (EFT) and the time-independent Schr\"odinger equation is then solved using nonperturbative many-body methods.

Chiral EFT offers a systematic expansion of nuclear and electroweak forces involved in this decay, while encoding neglected higher-energy information in contact terms, whose low-energy constants (LECs) are fit to reproduce chosen experimental data~\cite{Epelbaum2009, Machleidt2011}. 
In this work, the valence-space in-medium similarity renormalization group (VS-IMSRG)~\cite{Bogner2014,Stroberg2017,Stro19ARNPS,Miya20lMS} is used to approximately solve the Schr\"odinger equation and obtain NMEs from different starting chiral EFT potentials. 
This method decouples an effective valence-space Hamiltonian from the full space via continuous unitary transformations, therefore (in the absence of many-body truncations) preserving the eigenstates of the original Hamiltonian, while reducing significantly the size of the space considered. 
The valence-space Hamiltonian is then diagonalized using the shell-model code KSHELL~\cite{Shimizu2019}, thereby extending the reach of ab initio calculations to that of the traditional shell model~\cite{Stro21Drip}. 
Operators are then evolved using the same unitary transformations, allowing them to be treated consistently with the Hamiltonian. 
To make the problem computationally tractable, all operators are truncated at the normal-ordered two-body level, introducing the primary many-body approximation, IMSRG(2), in the NMEs. 
We further restrict the size of the initial single-particle space following the $e_{\textrm{max}}$ truncation, which limits the possible state to those having $e = 2n+l \leq e_{\textrm{max}}$ where $n$ is the principal quantum number and $l$ is the orbital angular momentum.
Finally, 3N forces are truncated via $e_1+e_2+e_3 \leq E_{\textrm{3max}}$ to satisfy computational memory limitations. 
Until recently, the $E_{\textrm{3max}}$ truncation proved to be the key bottleneck for ab initio methods to reach the heavy isotopes of experimental relevance for $0\nu\beta\beta$. 
Fortunately, recent advances~\cite{Miyagi2022} have overcome this limitation, allowing for the first ab initio NMEs for $^{130}$Te and $^{136}$Xe~\cite{Belley_inprep}, two of the most predominant candidates for experimental searches, joining previous calculations for $^{48}$Ca~\cite{Yao20Ca48,Belley2021,Nova21Ca48}, as well as $^{76}$Ge, and $^{82}$Se~\cite{Belley2021}.

\section{\label{sec2:NME}Nuclear matrix element}
The NME can be separated into two parts, a long-range part which itself consists of Gamow-Teller (GT), Fermi (F), and Tensor (T) components~\cite{Engel2017} as well as a short-range part in the form of a contact (CT) operator~\cite{Ciri18CT}:
\begin{equation}
    \label{eq:NME}
    M^{0\nu\beta\beta} = M^{0\nu\beta\beta}_{GT} - \big(\dfrac{g_V}{g_A}\big)^2 M^{0\nu\beta\beta}_{F}  + M^{0\nu\beta\beta}_T - 2 g_{NN}M^{0\nu\beta\beta}_{CT}.
\end{equation}
The contact operator, in particular, has been discovered recently to be promoted to leading order when considering the mechanism responsible for $0\nu\beta\beta$ at the electroweak scale within an EFT framework~\cite{Ciri18CT}. 
Due to this recent discovery, most calculations have not considered this important term until now~\cite{Wirr21CT}. 
Furthermore, the coupling constant, $g_{NN}$, is an unknown that needs to be fit to some lepton-number-violating observable, which of course do not exist. 
Nonetheless, a set of synthetic data of two-body amplitudes has been created by the same group that discovered the new term, allowing an estimation of the coupling constant to an accuracy of $\sim$30\%, when fitting to this data for a specific interaction~\cite{Cirigliano2021,Ciri21CT}. 
As this approach requires knowledge of a particular NN force, it is only fully consistent with ab initio methods, reinforcing the need for such calculation of the NME.
Finally, we have found a fairly large interaction dependence when considering several of the most commonly used chiral EFT interaction in the field~\cite{Simo17SatFinNuc, Jiang2020,Leistenschneider2018}.  
This calls for a more systematic way to constraint the uncertainty of the effect of the LECs on the NME.

\section{\label{sec3:uncertainty} Constraining the uncertainty}

We now look into further ways to study the interaction uncertainty. 
One approach is to consider the correlation between $M^{0\nu\beta\beta}$ and the matrix elements of other observables in different isotopes. 
One observable which has been suggested by phenomenology to be correlated is the double Gamow-Teller (DGT) transition, a charge-exchange reaction~\cite{Shimizu:2018PRL,Brase2022} predicted by the standard model but which remains to be observed. 
We have studied this correlation from ab initio methods in Ref.~\cite{Yao2022} and found that, even though a similar correlation can be extracted, it becomes significantly weaker when considering isospin-changing transitions, as are all experimental candidates for $0\nu\beta\beta$ decay. 
Due to the weak correlation in the experimentally relevant isotopes, the constraints that might have been put on the final $0\nu\beta\beta$-decay NME from this correlation, in a hypothetical observation of DGT, would be weaker than the current constraints coming from the interaction uncertainties.

A different approach would be to study the interaction uncertainty in a more systematic way, through studying the effect of varying the LECs in chiral forces. 
Doing this in the traditional way (i.e., by doing full ab initio calculation for each set of LECs) is extremely computationally expansive and therefore unfeasible at this time. 
However, due to the development of emulators based on eigenvector continuation~\cite{Frame2018, Ekstrtom2019}, it is now possible to probe the dependence of the results on the LECs, then use statistical tools such as history matching to select non-implausible sets of LECs that adequately describe chosen data. 
Such an analysis was carried out in~\cite{Hu2021}, where a set of 34 interactions based on the Delta-full chiral EFT~\cite{Jiang2020} (a theory which explicitly consider delta isobars) were extracted. 
From the spread du these interactions, we can explore the dependence of the $0\nu\beta\beta$-decay NME on the LECs. 
Furthermore, we can study the correlation with different observables in order to constrain the relevant LECs for this process.  

\begin{figure}
    \centering
    \includegraphics[width=\textwidth]{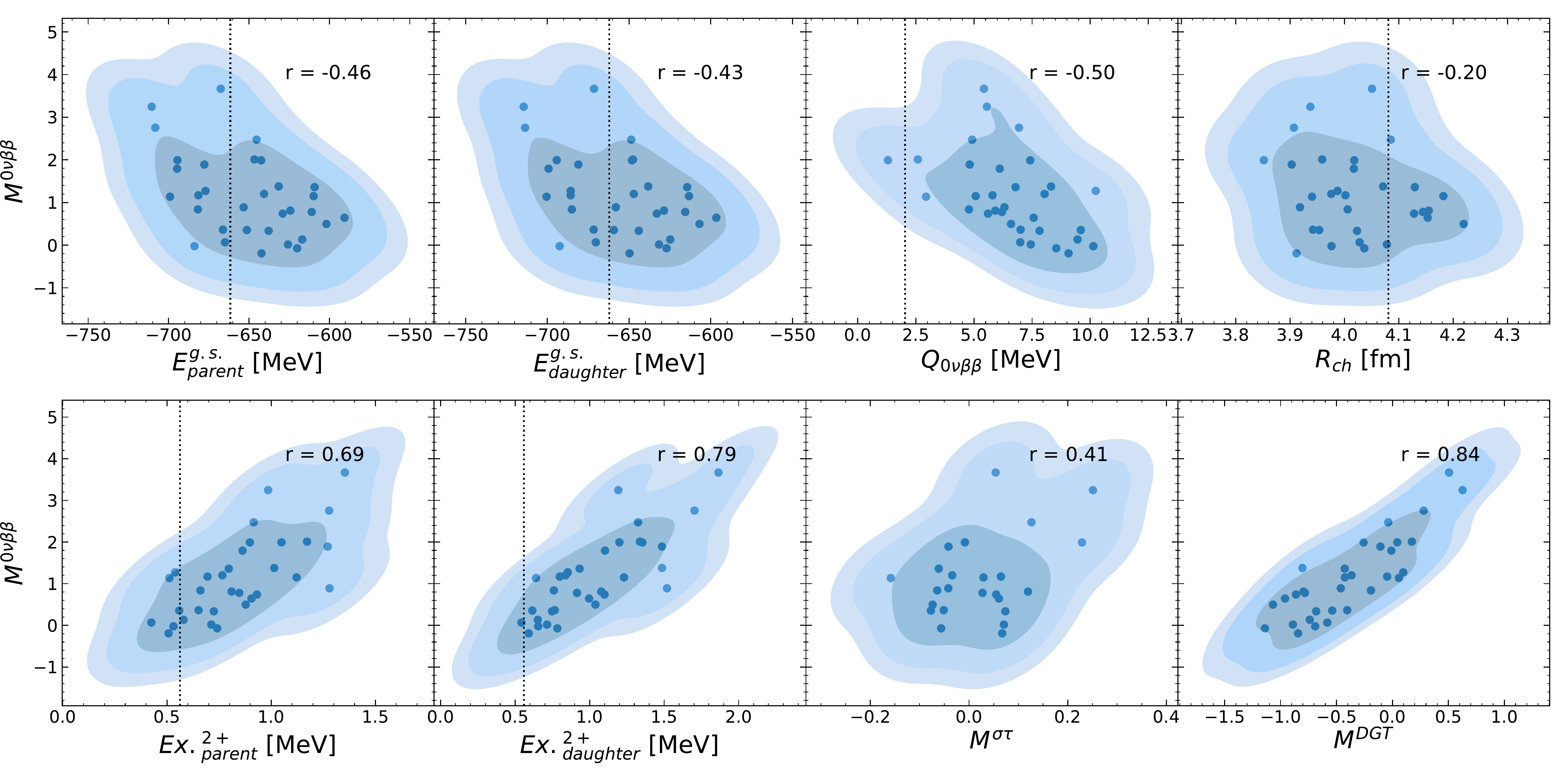}
    \caption{Correlation between different observables in $^{76}$Ge using the 34 non-implausible samples of parameters of N$^2$LO delta-full chiral EFT with a cutoff of 394 MeV\cite{Hu2021}. Results are compared to experiment~\cite{Wang2017,Angel2013} (dashed black lines) where available. The Pearson-R coefficient is given on each correlation plot to indicate the level of correlation between any two observables. The shaded region shows the probability density of finding the data points inside this region for 1$\sigma$, 2$\sigma$ and 3$\sigma$ respectively.
    }
    \label{fig:correlation}
\end{figure}

Figure~\ref{fig:correlation} shows the correlations for the transition $^{76}$Ge $\rightarrow ^{76}$Se between the ground-state energies of the parent/daughter ($E^{g.s.}_{parent}$/$E^{g.s.}_{daughter}$), the Q-value ($Q_{0\nu\beta\beta}$), the 2$^+_1$ energies of the parent/daughter ($Ex.^{2+}_{parent}$/$Ex.^{2+}_{parent}$), the charge radius of $^{76}$Ge ($R_{ch}$), the NME of the single beta decay from the ground state of $^{76}$Ge to the first excited state of $^{76}$As ($M^{\sigma\tau}$), $M^{DGT}$, and $M^{0\nu\beta\beta}$. 
We first note the vast spread of values of $M^{0\nu\beta\beta}$, indicating the LECs relevant for this process do not appear to be well constrained at the moment, at least at N$^2$LO in chiral EFT. 
To help constrain the NMEs, we look at other operators that might correlate with $M^{0\nu\beta\beta}$ decay. 
We find that the only viable candidates would be the 2$^+$ energies and $M^{DGT}$. 
The correlation with the 2$^+$ energies, however, is not seen outside of $^{76}$Ge, leaving only $M^{DGT}$ for the isotopes of interest. 
This is likely due to the larger deformation of the nuclei involved in the $^{76}$Ge decay. 
Therefore, even if the DGT transition may not help constrain the uncertainty via a nuclei-wide correlation, it might help constrain the LECs relevant to $0\nu\beta\beta$ decay, therefore reducing the interaction uncertainty.

\section{Conclusion}

In this work, we present ab initio results for the NMEs of $^{76}$Ge to analyze the interaction uncertainty by making use of recent novel advances in the field. 
We find that the LECs relevant to this process are generally not well constrained, but that observation of the DGT transition could potentially help reduce the interaction uncertainty on the NMEs.

\begin{acknowledgments}
We thank L. Jokiniemi for enlighting discussions and A. Ekström, C. Forssén, G. Hagen, and W. G. Jiang for providing the 34 sample interactions used in this work. TRIUMF receives funding via a contribution through the National Research Council of Canada. This work was further supported by NSERC under grants SAPIN-2018-00027 and RGPAS-2018-522453, the Arthur B. McDonald Canadian Astroparticle Physics Research Institute, and the Deutsche Forschungsgemeinschaft (DFG, German Research Foundation) -- Project-ID 279384907 -- SFB 1245.
Computations were performed with an allocation of computing resources on Cedar at WestGrid and Compute Canada.
\end{acknowledgments}

\nocite{*}
\bibliography{aipsamp}

\end{document}